\documentclass[prx,twocolumn,floatfix,nopacs]{revtex4-1}
\usepackage{graphicx,xcolor,amsfonts,amssymb,amsmath,enumerate,upgreek,tabularx,multirow}
\definecolor{link}{RGB}{46,46,145} 
\usepackage[colorlinks=true,allcolors=link]{hyperref}
\usepackage[displaymath,mathlines]{lineno}
\usepackage{mwe}
\usepackage{color}

\newcommand{\subfig}[2]{%
	{\textsf{#1}} \vtop{%
		\vskip0pt
		\hbox{#2}
	}}

\newcommand{\comment}[1]{}

\newcommand*{\bra}[1]{\mathopen{\langle}#1\mathclose{|}}
\newcommand*{\ket}[1]{\mathopen{|}#1\mathclose{\rangle}}
\newcommand{\ketbra}[1]{\ket{#1}\hspace{-0.25em}\bra{#1}}
\newcommand{\ketbrap}[2]{\ket{#1}\hspace{-0.25em}\bra{#2}}

\newcommand{\norm}[1]{\left\lVert#1\right\rVert}
\begin{document}

\title{Unpredictability and entanglement in open quantum systems}

\author{Javad Kazemi}
\email{javad.kazemi@itp.uni-hannover.de}
\affiliation{Institut f\"ur Theoretische Physik, Leibniz Universit\"at Hannover, Appelstra{\ss}e 2, 30167 Hannover, Germany}
\author{Hendrik Weimer}
\email{hweimer@itp.uni-hannover.de}
\affiliation{Institut f\"ur Theoretische Physik, Leibniz Universit\"at Hannover, Appelstra{\ss}e 2, 30167 Hannover, Germany}

\begin{abstract}

  We investigate dynamical many-body systems capable of universal
  computation, which leads to their properties being unpredictable
  unless the dynamics is simulated from the beginning to the
  end. Unpredictable behavior can be quantitatively assessed in terms
  of a data compression of the states occurring during the time
  evolution, which is closely related to their Kolmogorov
  complexity. We analyze a master equation embedding of classical
  cellular automata and demonstrate the existence of a phase
  transition between predictable and unpredictable behavior as a
  function of the random noise introduced by the embedding. We then
  turn to have this dynamics competing with a second process inducing
  quantum fluctuations and dissipatively driving the system to a
  highly entangled steady state. Strikingly, for intermediate strength
  of the quantum fluctuations, we find that both unpredictability and
  quantum entanglement can coexist even in the long time
  limit. Finally, we show that the required many-body interactions for
  the cellular automaton embedding can be efficiently realized within
  a variational quantum simulator platform based on ultracold Rydberg
  atoms with high fidelity.

\end{abstract}

\maketitle

\section{Introduction}

The discovery of problems that are fundamentally undecidable is one of
the most striking results in the history of mathematics
\cite{Godel1931,Turing1937}. For physical systems, this means that
certain problems such as the existence of a spectral gap in a
many-body system can be undecidable as well
\cite{lloyd1993,Cubitt2015}. Likewise, the long-term dynamics of a
physical system can be \emph{unpredictable} in the sense that the only
way to compute observables is to simulate the system from the
beginning to the end \cite{Wolfram1984}. Here, we show that
driven-dissipative quantum systems provide an ideal platform to study
such fascinating systems and to explore the hitherto largely unknown
relation of this unpredictably with entanglement that is absent in
classical dynamical systems. Strikingly, while we find that the
statistical aspects of entangled states inherently drive the system
towards chaotic behavior, we find that under the right conditions,
unpredictability and entanglement can coexist in the steady state.

The dynamical behavior of unpredictable systems is most readily
analyzed in the framework of cellular automata, as one can show
rigorously that these systems are Turing-complete and hence
unpredictable
\cite{Margolus1984,Berlekamp2004,Cook2004,Neary2006}. However, common
quantum version of cellular automata
\cite{Brennen2003,Raussendorf2005a,Arrighi2008,Bleh2012,Hillberry2021}
rely on purely unitary dynamics and are therefore challenging to
realize without having access to fault-tolerant quantum computers. For
this reason, variants of quantum cellular automata using open quantum
many-body systems \cite{Lesanovsky2019,Wintermantel2020} appear to be
a much more promising route to investigate the interplay between
unpredictability and inherently quantum properties of the dynamics
like quantum superpositions and entanglement.

\begin{figure*}[t]
	\begin{center}
		\begin{tabular}{ccc}
			& \subfig{a}{\includegraphics[width=.4\linewidth]
				{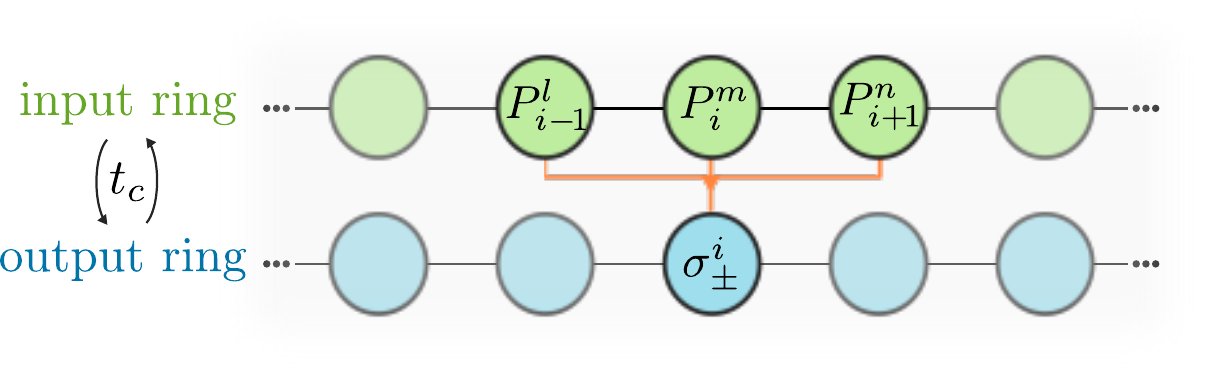}} &
			\\
			\subfig{b}{\includegraphics[height=.24\textheight] {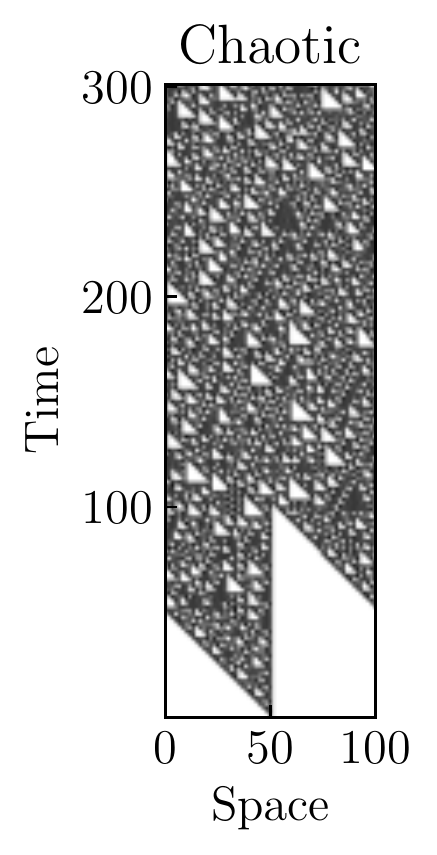}}
			&
			\subfig{c}{\includegraphics[width=0.4\linewidth]
			{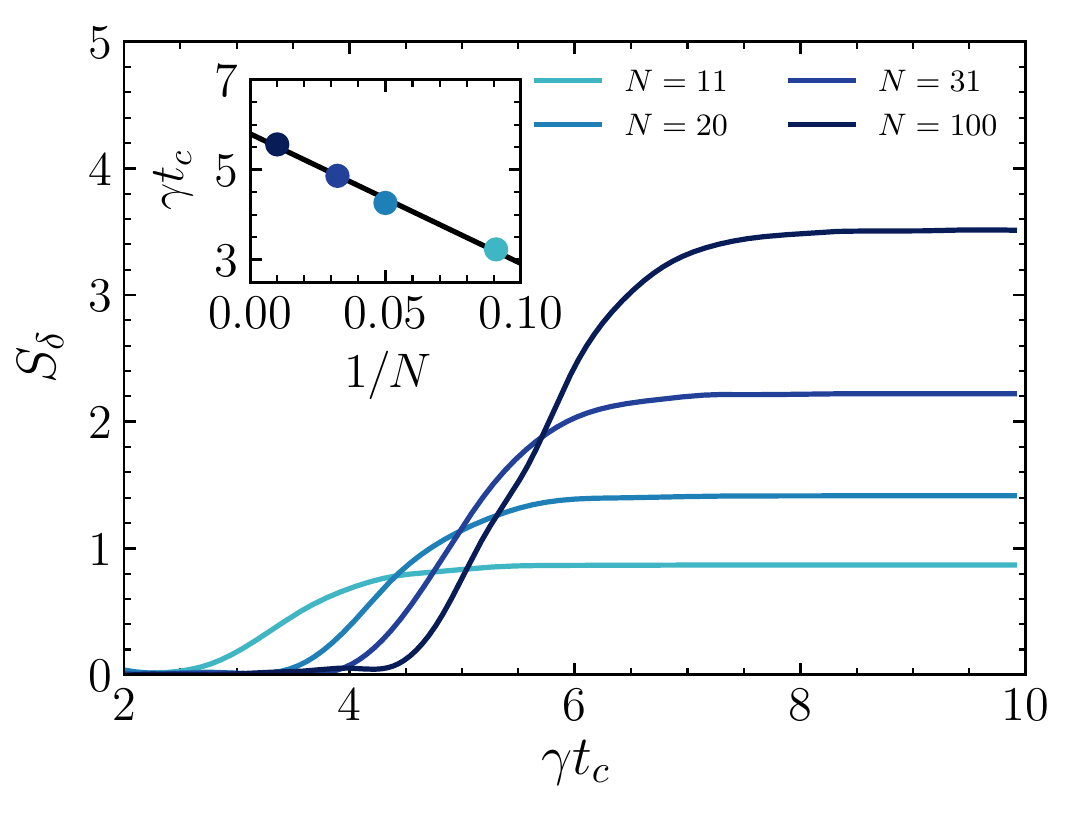}}
			&
			\subfig{d}{\vspace{10mm}\includegraphics[height=.24\textheight]
				{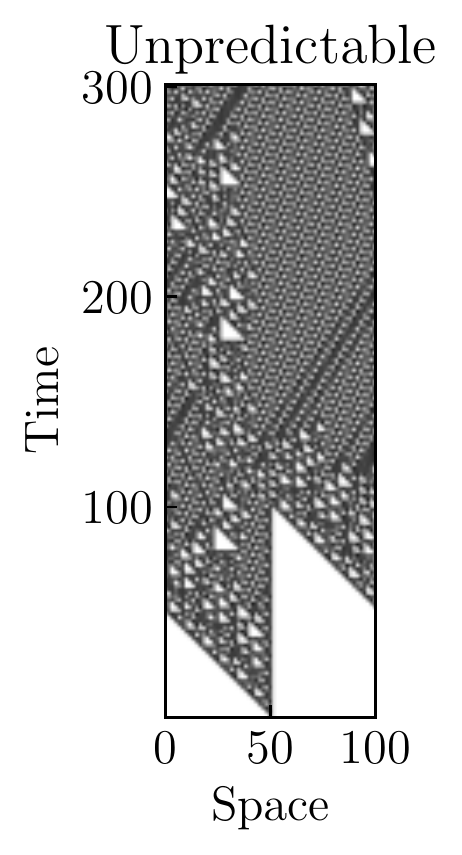}}\vspace{-12mm}
		\end{tabular}
	\end{center}
	\caption{(a) Physical implementation of elementary CA on two
          rings of lattice sites, with the input and output ring
          switching after a cycle time $t_c$. (b) Typical trajectory
          of the rule 110 CA evolving from a single cell in the 1
          state (black) for a short cycle time $t_c = 4/\gamma$,
          leading to Class III chaotic behavior. (c) Increasing the
          cycle time induces a transition to unpredictable behavior as
          shown by the slope $S_\delta$ of the characteristic
          compression exponent. Finite size scaling (inset) points to
          a value of $\gamma t_c = 5.80 \pm 0.08$ for the transition
          in the thermodynamic limit. (d) Typical trajectory in the
          unpredictable Class IV phase at a large cycle time $t_c
          = 10/\gamma$. At longer times, emergent glider structures
          propagating through the systems become clearly
          visible. Calculations were made at $n=50$ and $t=300$ to
          obtain convergence in $S_\delta$. For each initial state,
          100 Monte-Carlo samples were taken.}
	\label{fig:Sc}	
\end{figure*}

In this article, we provide an embedding of cellular automata (CA)
into a Lindblad master equation, which is commonly used within open
quantum systems. To this end, we consider two coupled one-dimensional
(1D) chains representing the state of the automaton at the time step
$t$ and $t+1$, respectively, see Fig.~\ref{fig:Sc}a. Here, we focus on
a class of elementary cellular automata introduced by Wolfram
\cite{Wolfram1983}, as this class contains Turing-complete automata
and can be realized in the embedding using at most four-body
interactions. The mapping of CAs onto master equation requires a
periodic switching of the chains, with the exact CA dynamics being
recovered in the limit of infinite cycle times. Interestingly, the
imperfections to the perfect CA rules introduced by finite cycle times
do not immediately destroy unpredictability, as shown using an
information-theoretical complexity measure based on data compression
of the states during the dynamical evolution of the system
\cite{Zenil2010}. We then introduce quantum fluctuations by adding a
second set of terms to the master equation that drive the system to a
highly-entangled Rokhsar-Kivelson state \cite{Roghani2018}. We explore
the interplay between unpredictability and entanglement by modifying
the relative rates of the two competing dynamics, finding an
intermediate regime in which both entanglement and unpredictability
coexist. Finally, we demonstrate the experimental feasibility of our
approach by investigating a platform based on ultracold Rydberg atoms
\cite{Browaeys2020,Morgado2021}. To this end, we introduce an open
system version of a variational quantum estimation algorithm
\cite{Peruzzo2014,Moll2018,Kokail2019} and show that the full dynamics
including dissipative many-body interactions can be realized
efficiently.

\section{Complexity classes and unpredictability}

The fundamental properties of the dynamical evolution of translation
invariant dynamical systems like CAs can be broadly captured in four
distinct classes \cite{Wolfram1984,Martinez2013}, referring to their
generic behavior for almost all initial states. Class I refers to a
fast evolution towards a steady state. Class II systems evolve towards
periodic limit-cycle oscillations. Class III systems are chaotic,
which can be captured in terms of a positive Lyapunov exponent
\cite{Eckmann1985} and fast relaxation of entropic measures to a
constant value that is large. Finally, Class IV systems exhibit
diverging transient times and are exhibiting emerging structures that
can be very complex. It is conjectured that Class IV systems are
capable of universal computation and hence their behavior in the
thermodynamical limit are fundamentally uncomputable
\cite{Wolfram1984}. It has also been discussed that Class IV systems
are critical phases at the edge between regular periodic and chaotic
behavior \cite{Langton1990}.

While chaotic (i.e., Class III) systems are sometimes also classified
as unpredictable because of their sensitvity to changes in the initial
condition, the notion of unpredictability we employ here is more
strict, as statements about \emph{some} long-time properties of
chaotic systems are possible in terms of statistical averages, e.g.,
by coarse-graining the system under consideration
\cite{Cotler2017}. Our notion of unpredictability is closely connected
to the concept of \emph{sophistication} in computer science
\cite{Koppel1987}, which can also be understood as capturing the
complexity of systems at a coarse-grained level \cite{Aaronson2014}.

The quantitative identification of Class IV behavior is extremely
challenging, requiring the computation of observables that go beyond
entropic measures or Lyapunov exponents geared to the analysis of
Class III systems. For systems of discrete variables, one possibility
is an information-theoretical analysis of the computational strings
encoding the state of the system. For each string $S$, the algorithmic
complexity is given by its Kolmogorov complexity $\mathcal{K}(S)$,
which is defined as the length of the shortest program that can output $S$ \cite{Li2008}. However, Kolmogorov
complexity itself is an uncomputable quantity, preventing a direct
practical application. Fortunately, it is possible to construct
practically useful upper bounds to the Kolmogorov complexity by
considering the compression length $\mathcal{C}(S)$ satisfying
\begin{equation}
  \mathcal{C}(S) = \mathcal{C}_c(S) + \mathcal{C}_0 \geq \mathcal{K}(S),
\end{equation}
where $\mathcal{C}_c(S)$ is the compressed length of $S$ using a
compression program of length $\mathcal{C}_0$, the latter being the
same for all strings $S$.

For the classification of dynamical systems, the Kolmogorov complexity
(or the compression length is not sufficient to capture
unpredictability, as the (pseudo-)randomness inherent in chaotic
systems will generically lead to large values of $\mathcal{K}(S)$ even
for Class III systems. However, it is possible to differentiate Class
III and Class IV by looking at the compression length for different
initial states \cite{Zenil2010,Martinez2013}. For chaotic Class III
systems, the Kolmogorov complexity does not depend on the particular
initial state, i.e., differential measures of Kolmogorov complexity
are vanishing. Importantly, this is not the case in Class IV systems, where
the Kolmogorov complexity strongly depends on the initial state. This
can be understood as the initial state encoding a program that is run
using the universal computing capabilities of the Class IV
system. Some programs can produce simple computing results exhibiting
a low Kolmogorov complexity, while others can lead to arbitrary
complex computations. This can be captured quantitatively in the form
of a characteristic compression exponent \cite{Zenil2010}, which is a
generalization of the Lyapunov exponent to algorithmic complexity. It is given by
\begin{equation}
\delta_n(t) = \sum_{j=1}^{n-1}\frac{\big|\mathcal{C}_c(S_{j}(t)) -  \mathcal{C}_c(S_{j+1}(t)) \big|}{n-1},
\end{equation}
where $S_j(t)$ refers to the string representation of the state at
time $t$ for the $j$th initial state and the sum runs over $n$ initial
states in total. For systems with binary degrees of freedom, this can
be achieved by using the Gray code \cite{Gray1953}, which ensures that
the changes observed in the compression lengths are not stemming from
discontinuities within the initial conditions \cite{Zenil2010}. For
the compression algorithm, we use the \texttt{zlib.compress} function
provided by Python 3.6.9., which provides a widely used implementation
of the \textsc{DEFLATE} algorithm \cite{Deutsch1996}.

Generically, $\delta_n(t)$ will increase linearly with time in the
long time limit. Therefore, the slope $S_\delta = d \delta_n(t)/dt$ of the characteristic
exponent for sufficiently large $n$ is the actual quantity that can be
used to identify Class IV systems.

\section{Embedding of elementary cellular automata}

Elementary cellular automata (ECAs) \cite{Wolfram1984a} are two-level
systems on a 1D lattice, where the state $i$ at the dimensionless time
$t+1$ depends only on the states of the sites $i-1$, $i$, and $i+1$
and time $t$. In total, there are $2^{2^3} = 256$ different possible
ECA rules, which can be enumerated according to the binary
representation of their ruleset. Here, we will be interested in rule
110 (or its binary complement 137), see Tab.~\ref{tab:110} for the
ruleset, as this ECA has been proven to be capable of universal
computation \cite{Cook2004} and hence belongs into Class IV.
\begin{table}[b]

  \begin{tabular}{lcccccccc}\hline \hline
    State at $t$ & 111 & 110 & 101 & 100 & 011 & 010 & 001 & 000\\\hline
    Central site at $t+1$ & 0 & 1 & 1 & 0 & 1 & 1 & 1 & 0\\\hline\hline
  \end{tabular}
   
  \caption{Ruleset for ECA rule $110 = 01101110_2$ indicating how the central site is updated. The ruleset for the corresponding
    rule 137 can be found by inverting all inputs and outputs.}
\label{tab:110}
\end{table}

Most ECA rules are irreversible, as each output bit depends on three
input bits. This prevents an implementation in a quantum system in
terms of unitary operations and instead requires the use of a
dissipative quantum channel $\mathcal{V}$ providing the mapping
$\rho(t+1)=\mathcal{V}\rho(t)$ for the quantum state $\rho$. The
generator of such a quantum channel can be expressed in terms of a
purely dissipative quantum master equation in Lindblad form,
\begin{equation}
  \frac{d}{dt}\rho = \gamma\sum\limits_{i} \left(c_i\rho c_i^\dagger - \frac{1}{2}\left\{c_i^\dagger c_i, \rho\right\}\right),
\label{eq:master}
\end{equation}
with $c_i$ being the associated quantum jump operators
\cite{Breuer2002} and $\gamma$ being the characteristic rate of the
dynamics. Note that the channel $\mathcal{V}$ is only implemented in
the steady state, i.e., in the infinite time limit. For the case of a
finite evolution time $t_c$, this means that the evolution of the
corresponding cellular automaton becomes probabilistic. However, since
a finite error rate can be recovered in classical systems by error
correction codes, one can expect that Class IV systems capable of
universal computation can tolerate a finite amount of computational
errors introduced by the probabilistic update scheme. Interestingly,
this setting allows to treat the cycle time $t_c$ as a control
parameter, which allows to alter the dynamical properties of the
system under consideration in a relatively simple way.

For practical implementation purposes, it is highly desirable that the
quantum jump operators are quasi-local, i.e., they act only on a small
subset of the total Hilbert space. Within the ECA framework, this can
be readily achieved by considering a second copy of the system, in
which the state at time $t+1$ is prepared, see
Fig.~\ref{fig:Sc}a. After a time $t_c$, the role of the two copies is
reversed and the second copy is taken as the input for the creation of
the new state at time $t+2$ in the first copy.

\section{Cycle time phase transition}

\subsection{Purely classical dynamics}

\label{sec:classical}

Let us now turn to investigating how the dynamics of the rule 110 CA
change with the cycle time $t_c$. Since the dynamics is purely
classical, the master equation can be simulated efficiently using
Monte-Carlo sampling. Fig.~\ref{fig:Sc}c shows that the slope of the
characteristic compression exponent $S_\delta$ indeed undergoes a
phase transition from a chaotic Class III phase (Fig.~\ref{fig:Sc}b)
with vanishing $S_\delta$ to an unpredictable Class IV phase
(Fig.~\ref{fig:Sc}d) with finite $S_\delta$. We also observe that the
value of $S_\delta$ significantly depends on the system size $N$,
which can be attributed to the appearance of emergent glider
structures much larger than the three-site unit cell of the CA
\cite{Cook2004}. For example, the smallest system size in which all
types of gliders can be realized is $N=11$. Nevertheless, we can still
perform finite size scaling of $S_\delta$ by considering only those
system sizes at which $S_\delta$ is at a local maximum, a technique
that has been used previously in the analysis of systems with strong
incommensurability effects \cite{Weimer2010a,Arora2015}. Here, we find
the phase transition to occur at $t_c = 5.80 \pm 0.08$
 in the thermodynamic limit. Note that it
might seem surprising that we are able to perform finite size scaling
in a system with unpredictable behavior, as unpredictability is
directly tied to a breakdown of finite size scaling theory
\cite{Cubitt2015}. However, here one of the two phases has the
computable value of zero for $S_\delta$ in the thermodynamic limit,
which means that the phase transition can successfully detected by
observing deviations from $S_\delta$ vanishing for increasing system
sizes.

\subsection{Competition with entanglement}

So far, we have not addressed including quantum fluctuations into the
dynamics. While there are many different proposals for CA dynamics
within quantum systems
\cite{Brennen2003,Bleh2012,Lesanovsky2019,Wintermantel2020,Klobas2020,Wilkinson2020},
here we focus on the case where we have two competing dynamics
$\mathcal{L}_c$ and $\mathcal{L}_q$, which are both generators
satisfying the purely dissipative Lindblad form of
Eq.~(\ref{eq:master}), i.e., the full Lindblad generator $\mathcal{L}$
is given by
\begin{equation}
  \mathcal{L} = \sin^2\left(\phi \frac{\pi}{2}\right)\mathcal{L}_c + \cos^2\left(\phi \frac{\pi}{2}\right) \mathcal{L}_q,
\end{equation}
where $\phi$ is a parameter that interpolates between the fully
quantum and fully classical dynamics. This approach has the advantage
that the relevant observable $S_\delta$ for detecting unpredictability
can be carried over to the quantum case by identifying the $i$th bit
in the string $S_j$ by the most likely measurement result when sampled
over many trajectories.

In the following, we will be interested in the case where the quantum
part of the dynamics is carefully chosen to have its unique steady
state being highly entangled. For this, we choose a recently
introduced quantum master equation that prepares the system in a
Rokhsar-Kivelson state \cite{Roghani2018}, which is given by
\begin{equation}
 |\psi_{RK}\rangle =\frac{1}{\sqrt{Z}}\prod_k^N(1- P_{k-1}\sigma_+^k P_{k+1})|00 \cdots 0\rangle,
\end{equation}
where $P_k = \ketbra{0}_k$ and $Z$ is a normalization constant. This
state is an equal-weight superposition of all states that have no
adjacent qubits in the $\ket{1}$ state. The jump operators $c_i^q$ required to
prepare $\ket{\psi_{RK}}$ can be brought into a similar form as the one for the rule 110, i.e.,
\begin{equation}
  c_i^q = \sum\limits_k O_{\bar{i},k} Q_{i,k}
\end{equation}
where $O_{\bar{i},k}$ is an operator acting on the sites $i-1$, $i$,
and $i+1$ on the \emph{output} ring, while $Q_{i,k} = \ketbra{k}$ is a
projection operator acting on the corresponding sites of the
\emph{input} ring. The index $k$ runs over all eight basis states on
the sites $i-1,$ $i$, and $i+1$. Table~\ref{tab:rk} shows a set of
operators that lead to $\ket{\psi_{RK}}$ being the dark state of the
dynamics, satisfying $c_i^q\ket{\psi_{RK}}=0$. To observe the
interplay between unpredictability and entanglement, it is helpful to
make the two competing dynamics closely aligned with each other. One
can see that $\ket{\psi_{RK}}$ will have more 0s than 1s in its binary
representation due to the action of the projection operators $P_k$. On
the other hand, the dynamics of rule 110 is slightly biased towards
the 1 state, as there are more 1s than 0s in the output
set. Therefore, we consider the binary complement rule $137 =
10001001_2$, which is then also biased towards the 0 state, but having
otherwise identical properties as rule 110.
\begin{table}[t]
	
	\begin{tabular}{lcccccccc}\hline \hline
    State $k$ at $t$ & 111 & 110 & 101 & 100 & 011 & 010 & 001 & 000\\\hline
    Operation $O_k$  & $\mu$ & $\sigma_-$ & $\mu$ & $\mu$ & $\mu$ & $\mu$ & $\mu$ & $\mu$\\\hline\hline
	\end{tabular}
	
	\caption{Ruleset to prepare the Rokhsar-Kivelson state
          $\ket{\psi_{RK}}$. The operator $\mu_{\bar{i}} =
          P_{\bar{i}-1} {\ketbrap{-}{+}_{\bar{i}}} P_{\bar{i}+1}$ acts
          on the three sites surrounding $\bar{i}$ on the output ring,
          with $\ket{\pm} = (\ket{0}\pm \ket{1})$. The operator
          $\sigma_-$ acts only on the site $\bar{i}$.}
\label{tab:rk}
\end{table}
To further align the two competing dynamics, we additionally split the
dynamics into two parts, each being $t_c/2$ long. During the first
part, both the classical part $\mathcal{L}_c$ and the quantum part
$\mathcal{L}_q$ are active, while during the second part only the
classical dynamics is acting on the system. All observables are
evaluated at the time $t_c/2$.

\begin{figure}[t]
	\begin{tabular}{c}
		\hspace{6mm}\subfig{a}{\includegraphics[width=.7\linewidth]{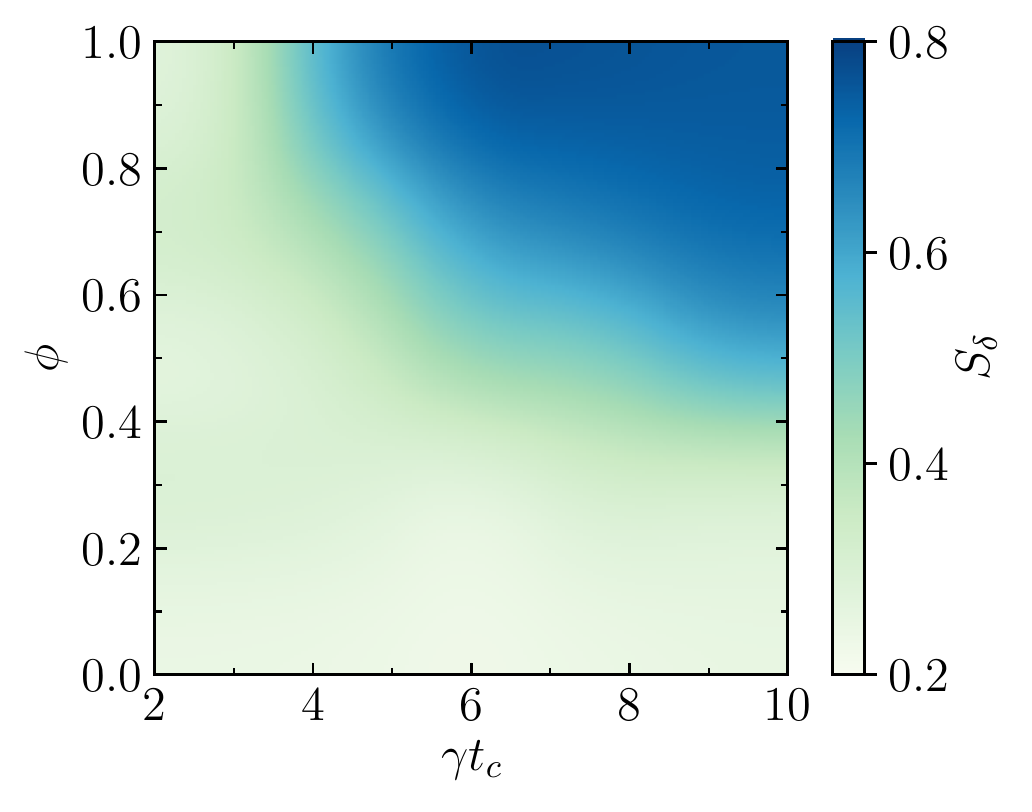}} \\
		\subfig{b}{\includegraphics[width=.9\linewidth]{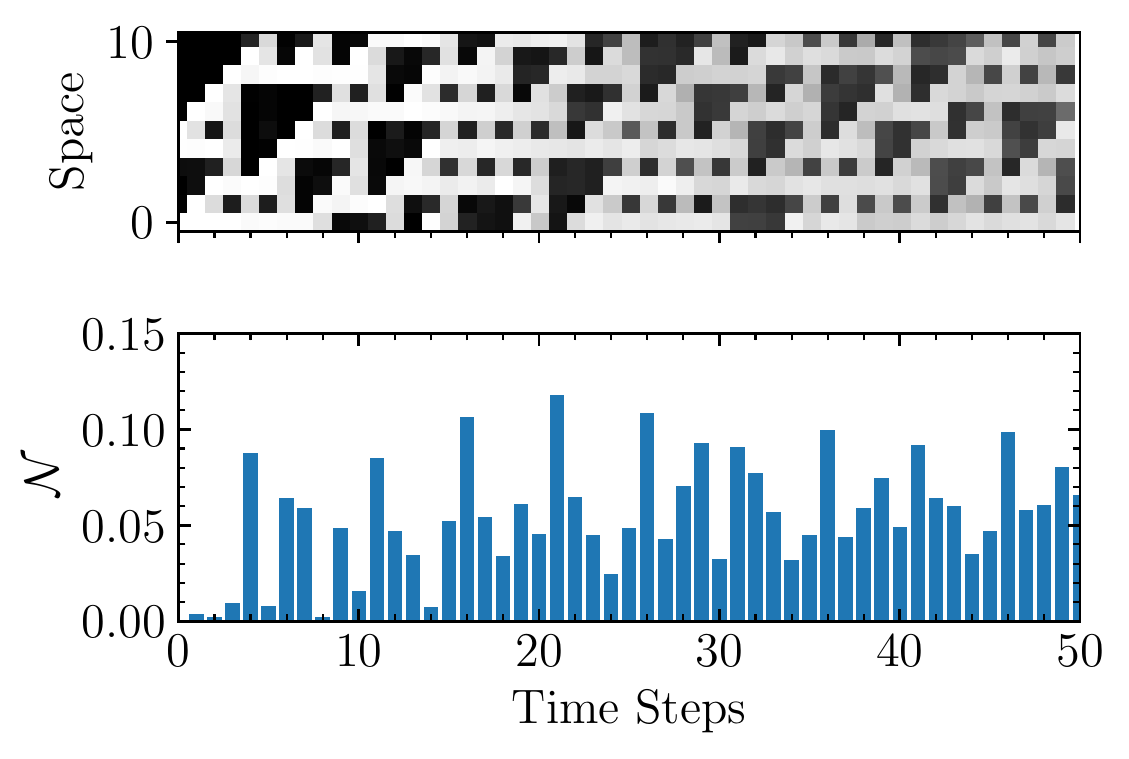}}
	\end{tabular}	
	\caption{Competition between classical and quantum
          dynamics. (a) Slope $S_\delta$ of the characteristic
          compression exponent as a function of the cycle time $t_c$
          and the interpolation parameter $\phi$, showing an extended
          region of unpredictable Class IV behavior even in the
          presence of quantum fluctuations. (b) Average over 50
          Monte-Carlo samples of a single initial configuration for
          $\gamma t_c = 10$ and $\phi = 0.6$. Although quantum
          fluctuations lead to a washing out of the patterns, the
          emergent structures remain clearly visible. The negativity
          $\mathcal{N}$ rapidly increases and remains finite in the
          long time limit.}
	\label{fig:Q137}	
\end{figure}
Let us now turn to a numerical investigation of the competition
between the classical rule 137 dynamics $\mathcal{L}_c$ and the
quantum dynamics $\mathcal{L}_q$ preparing the Rokhsar-Kivelson
state. Figure~\ref{fig:Q137} shows the unpredictability of the system
in terms of the slope $S_\delta$ of the characteristic compression
exponent as a function of the cycle time $t_c$ and the interpolation
parameter $\phi$, as well as the entanglement of the system in terms
of the negativity
\begin{equation}
  \mathcal{N}(\rho) = \frac{\norm{\rho^{T_A}}_1-1}{2},
\end{equation}
where $\norm{.}_1$ denotes the trace norm and $\rho^{T_A}$ is the
partial transpose of $\rho$ with respect to the subsystem $A$
\cite{Vidal2002}. Calculations were performed using massively
parallelized wave-function Monte-Carlo simulations
\cite{Johansson2013} for 22 sites, improving the previously reported
largest system size for simulations retaining the full Hilbert space
\cite{Raghunandan2018}. For the classical limit $\phi \to 1$, we
recover the classical transition to an unpredictable Class IV phase
reported in Sec.~\ref{sec:classical}. However, this transition remains
observable away from the classical limit as well. Crucially, we also
find a nonzero negativity for sufficiently low values of $\phi$,
meaning that unpredictability and entanglement can coexist in the
system. Finally, when the interpolation parameter $\phi$ is decreased
further, unpredictable Class IV behavior is destroyed by quantum
fluctuations.

\section{Variational quantum simulation}

The realization of the jump operators for the classical and quantum
dynamics requires the implementation of four and six-body
interactions, respectively. While such high-order interactions can be
readily engineered in digital quantum simulators
\cite{Weimer2010,Weimer2011}, recently developed techniques for the
variational preparation of quantum states
\cite{Peruzzo2014,Moll2018,Kokail2019} can be adapted for the quantum
simulation of open system dynamics, especially on noisy
intermediate-scale quantum devices. Basically, variational quantum
simulation (VQS) is a hybrid classical-quantum approach that can
approximately reproduce the dynamics of a quantum system without
explicit realization of the corresponding Hamiltonian.

Within our VQS approach, we can realize the time evolution of largely
arbitrary open quantum many-body systems according to a Lindblad
quantum master equation. To this end, we consider the dynamics of an
observable $O$ in the Heisenberg picture,
\begin{equation}\label{eq:Heisenberg}
\frac{d O }{dt} \equiv \mathcal{L}_HO = i [H,O] +\sum\limits_{i}^N  {L_i}^\dagger O L_i - \frac{1}{2} \big\{ L_i^\dagger L_i, O \big\}.
\end{equation}
The strategy behind VQS is to consider a variational parametrization
of the state $\rho(t+\tau)$ after a timestep $\tau$, given the state
$\rho(t)$ at time $t$ and the dynamics according to the quantum master
equation. Our approach is based on the variational principle for open
quantum systems \cite{Weimer2015}, which also has been generalized
towards variational approximation of the time evolution, both in the
Schr\"odinger \cite{Overbeck2016} and in the Heisenberg picture
\cite{Pistorius2020a}. Within the latter, one can introduce a
variational cost functional $F_v$ given by
\begin{equation}
  F_v = \sum_{i}\Big|\big\langle O_i(t+\tau)\big\rangle-\big\langle O_i(t)\big\rangle-\frac{\tau}{2} \mathcal{L}_{H}[O_i(t)+O_i(t+\tau)] \Big|,
  \label{eq:cost}
\end{equation}
which realizes a discretized version of the master equation that is
correct up to second order in $\tau$. Crucially, the sum does not need
to run over a complete set of observables to achieve accurate results
\cite{Pistorius2020a}.

\begin{figure}[t]
	\includegraphics[width=1\columnwidth]{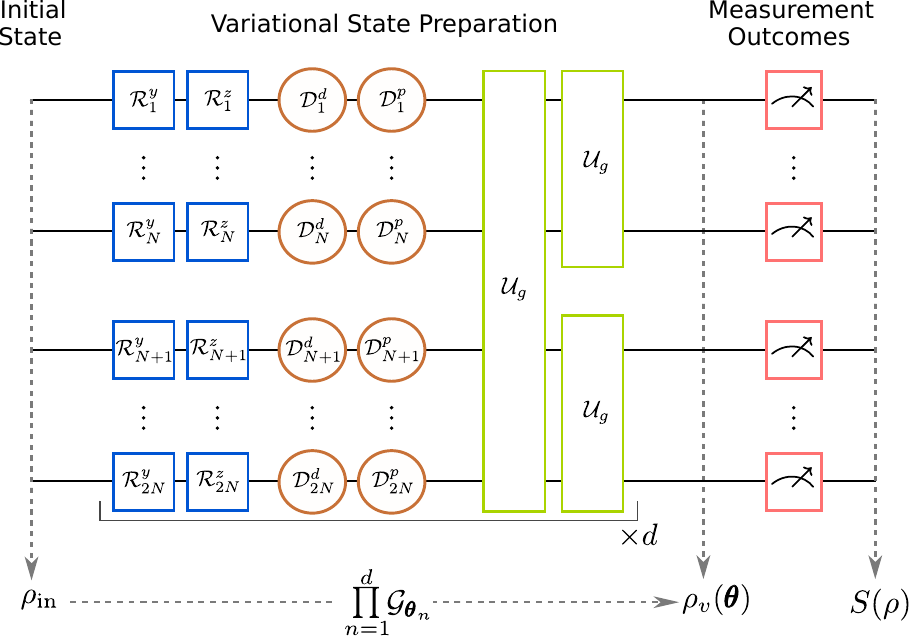}
	\caption{Sketch of the variational quantum simulation
          scheme. A quantum state $\rho_v(\boldsymbol{\theta})$ is
          generated from an initial state $\rho_{\text{in}}$ and a set
          of variational parameters
          $\boldsymbol{\theta}=\{\boldsymbol{\theta}_n\}_{n=1}^{d}$
          for $d$ layers. The variational circuit consists of local
          Pauli rotations $\mathcal{R}(\boldsymbol{\theta}_n)$, local
          dissipation operators $\mathcal{D}(\boldsymbol{\theta}_n)$,
          and global unitary operations
          $\mathcal{U}_g(\boldsymbol{\theta}_n)$.}
	\label{fig:QVC}	
\end{figure}

Having defined the variational cost function, we also need to specify
the parameterization of the density matrix $\rho$ using a set of
variational parameters $\boldsymbol{\theta}$. For this, we consider a
variational circuit consisting of three parts, see Fig.~\ref{fig:QVC}:
(i) Coherent single qubit rotations, (ii) dissipative single qubit
damping, and (iii) global unitary time evolution under an interacting
many-body Hamiltonian that can be implemented efficiently.

To be specific, we envision the qubits stored in the hyperfine ground
states of ultracold atoms trapped in an optical tweezer array
\cite{Endres2016,Barredo2016}. While the coherent single qubit
rotations are following from standard Pauli rotations around two
independent axes using microwave driving, the dissipative dynamics is
governed by two separate quantum channels realized by dissipative
optical pumping of the qubit states ($\mathcal{D}_d$) and phase
fluctuations mediated by noisy laser driving ($\mathcal{D}_p$),
respectively. In the operator sum representation \cite{Nielsen2000}
the two channels can be written as
\begin{equation}
\mathcal{D}_{\mu}(\rho) = \frac{1}{N} \sum_{j=1}^{N} \Big[ D_{\mu_1}^{(j)} \rho {D_{\mu_1}^{(j)}}^\dagger+D_{\mu_2}^{(j)} \rho {D_{\mu_2}^{(j)}}^\dagger \Big],
\end{equation}
where $\mu \in \{d,p\}$ and
\begin{align}
D_{d_1} =& \sqrt{1-e^{- \theta_d}} \sigma_{-},\\
D_{d_2} =& P_0 + e^{-\theta_d/2}P_1,\\
D_{p_1,p_2} =& \sqrt{\frac{e^{-\theta_p/2}}{2}}  \pm  \sqrt{\frac{(1-e^{-\theta_p})}{2}} \sigma_{z},
\end{align}
depending on the variational parameters $\theta_d$ and
$\theta_p$. Finally, the global unitaries $\mathcal{U}_g$ are
implemented based on coupling the $\ket{1}$ state to a strongly
interacting Rydberg state in a Rydberg dressing configuration
\cite{Glaetzle2015,Zeiher2016,Overbeck2017,Helmrich2018}, giving rise
to the effective Hamiltonian
\begin{equation}
H_{0} = \frac{\Omega}{2} \sum_{i} \sigma_x^{(i)} + \sum_{i<j} \frac{C_6}{{|\textbf{r}_i-\textbf{r}_j|}^6}P_1^{(i)}P_1^{(j)},
\end{equation}
where $\Omega$ indicates the Rabi frequency of the driving field and
$C_6$ denotes the effective van der Waals coefficient of the
Rydberg-dressed state. The global unitaries act first on both of the
rails at the same time and then subsequently on each rail
independently. The times $t_1$ and $t_2$ of these steps provide two
additional variational parameters.

\begin{figure}[t]
	\begin{tabular}{c}	
		\subfig{a}{\includegraphics[width=.8\linewidth]{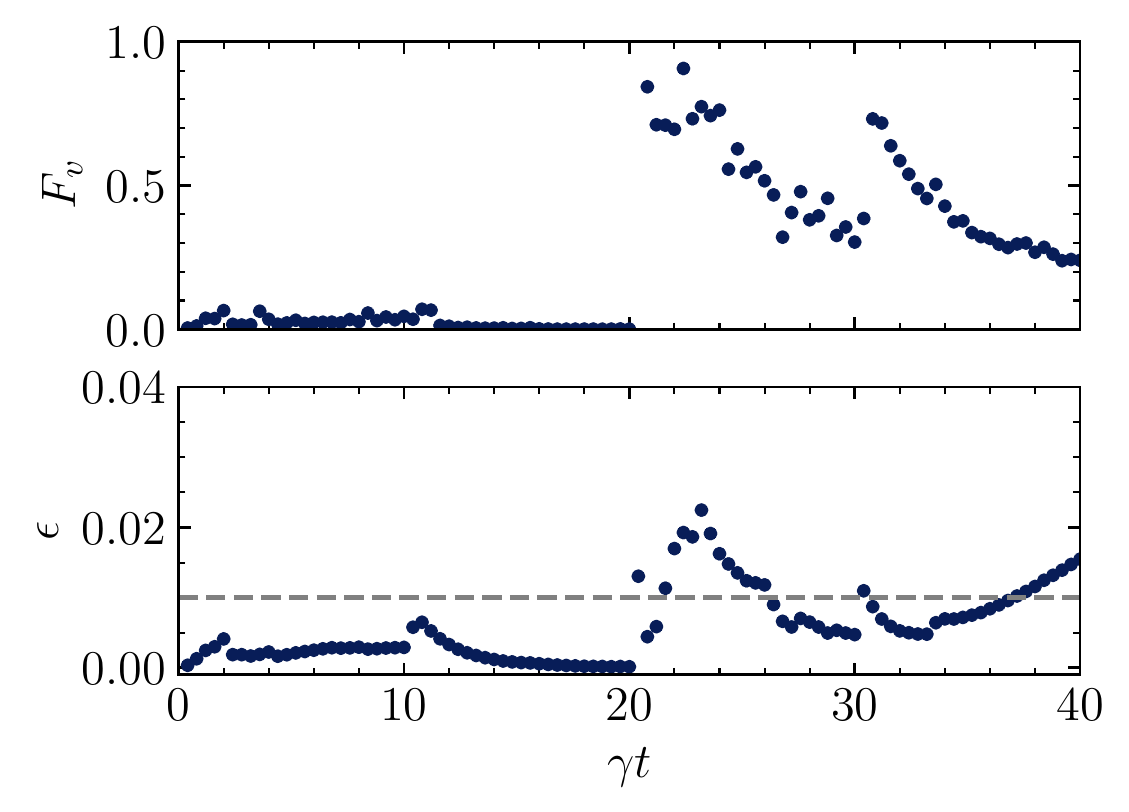}}\\
		\subfig{b}{\includegraphics[width=.8\linewidth]{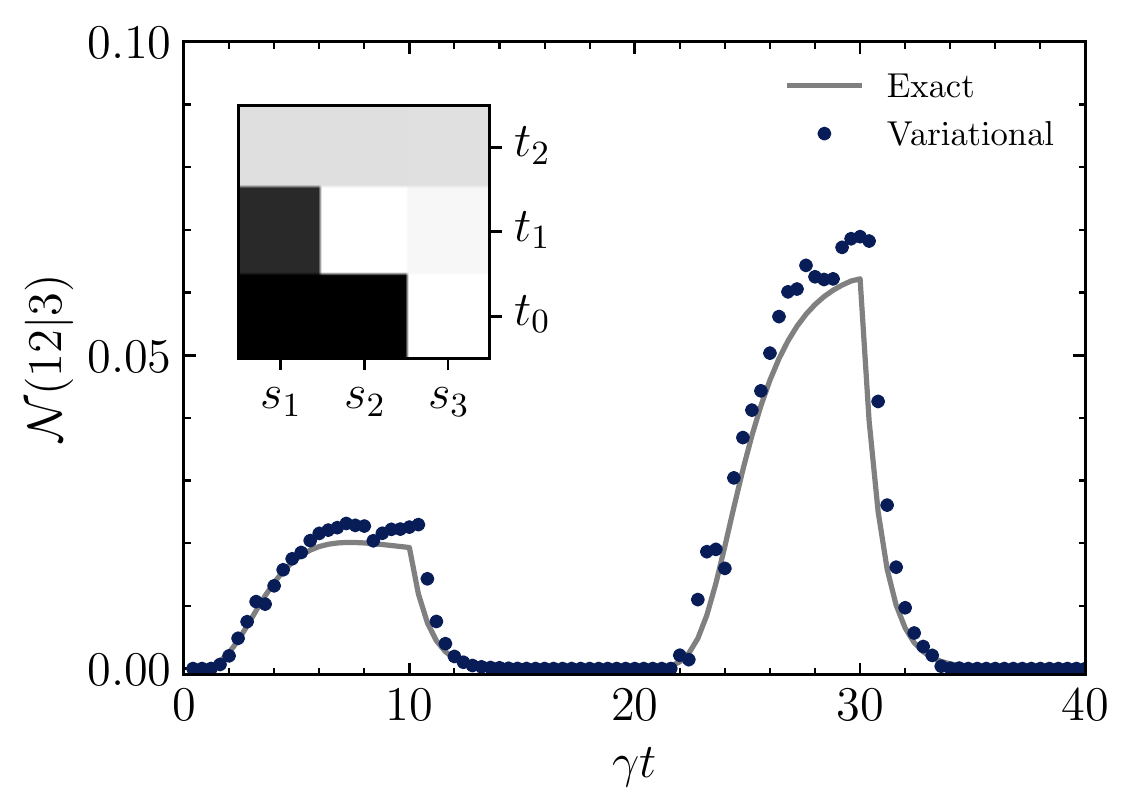}}
	\end{tabular}
	\caption{Benchmarking of the VQS approach for $\phi = 0.5$ and
          $\gamma t_c = 10$ for a system of $N=3$ sites, using $d=3$
          variational layers. Variational norm $F_v$, error
          probability $\varepsilon$ (a) and quantum negativity
          $\mathcal{N}$ (b) as a function of time for the first two
          timesteps. The inset shows the density pattern in the output
          ring at the middle of the steps.}
	\label{fig:QECA137}	
\end{figure}
Let us now benchmark our VQS procedure for the quantum rule 137
dynamics using two rings of $N=3$ sites. In our simulations, we set
$C_6/\Omega=100$ as this corresponds to a Rydberg blockade radius $r_b
= \sqrt{6}{C_6/\Omega}\approx 2.15$ in units of the lattice spacing,
which is comparable to the range of the multi-qubit jump operators in
the quantum master equation. Furthermore, we choose a variational
circuit depth of $d=3$. For the set of observables entering the
variational cost function (\ref{eq:cost}), we choose a complete set of
3-local Pauli operators per site, i.e., $4^3 = 64$
observables. Numerical optimization of the highly non-linear cost
function is performed using a sequential quadratic programming
algorithm \cite{Virtanen2020} in a layerwise fashion to avoid getting
stuck in barren plateaus \cite{McClean2018}.

Figure~\ref{fig:QECA137}a shows variational norm $F_v$ and error probability $\varepsilon$ between the exact density matrix $\rho$ and variational state $\rho_v(\boldsymbol{\theta})$ at time $t$, defined in terms of the fidelity $F$ as 
\begin{equation}
\varepsilon=1-F=1-\Bigg[\text{Tr}\sqrt{\sqrt{\rho_v(\boldsymbol{\theta})}\rho\sqrt{\rho_v(\boldsymbol{\theta})}}\Bigg]^2.
\end{equation}
For $\phi=0.5$ and $\gamma t_c = 10$, where unpredictability and
quantum entanglement coexist, the results indicate overall errors less
than $2\% $ for almost all data points. Figure~\ref{fig:QECA137}b
demonstrates that entanglement measured in terms of the negativity
also shows good quantitative agreement with the exact dynamics,
underlining the power of the VQS approach.

\section{Conclusion}

In summary, our work establishes a novel dynamical class of open
quantum many-body systems that allow to study the interplay between
computational properties and quantum effects. Strikingly, we have
shown that computational unpredictability is not incompatible with
quantum entanglement, but that the two can coexist with each other for
a long time. Our results on the variational quantum simulation of
these systems show that their experimental realization is possible
using present technologies, making these systems also excellent
candidates to observe a quantum advantage, given that their classical
simulation is extremely challenging \cite{Weimer2021}. In the future,
it will be interesting to develop quantum analogs of the data
compression approach for Class IV systems, using quantum compression
algorithms \cite{Schumacher1995,Jozsa1998,Rozema2014,Romero2017}.

\begin{acknowledgments}
The authors acknowledge fruitful discussion with R. van Bijnen on variational quantum simulation. This work was funded by the Volkswagen Foundation, by the Deutsche Forschungsgemeinschaft (DFG,German  Research  Foundation)  within  SFB  1227  (DQ-mat, project A04), SPP 1929 (GiRyd), and under Germanys  Excellence  Strategy  –  EXC-2123  QuantumFrontiers – 390837967.
\end{acknowledgments}

\bibliographystyle{myaps}
\bibliography{../bib/bib}


\end{document}